\documentclass[%
 aip,%
 apl,%
 amsmath,amssymb,%
 showkeys,%
 reprint,%
 floatfix,%
 superscriptaddress,
 nobalancelastpage
]{revtex4-1}

\usepackage[product-units = power,range-units = single]{siunitx}
\usepackage{graphicx}
\usepackage{graphicx}
\usepackage{verbatim}
\usepackage{epstopdf}
\usepackage{dcolumn}
\usepackage{bm}
\usepackage{siunitx}
\usepackage{lineno}
\usepackage{textcomp}
\usepackage{xr}
\usepackage{subfigure}

\DeclareSIUnit\BohrMagneton{$\mu_mathrm{B}$}
\DeclareSIUnit\formulaunit{f.u.}
\DeclareSIUnit\atomicunit{a.u.}
\DeclareSIUnit\arbunit{arb.unit}
\DeclareSIUnit\torr{Torr}
\usepackage{natbib}

\begin{document}
\title{Superconducting granular aluminum resonators resilient to magnetic fields \\ up to 1 Tesla}

\author{K.~Borisov}
\email[]{kiril.borisov@kit.edu}
\affiliation{IQMT,~Karlsruhe~Institute~of~Technology,~76344~Eggenstein-Leopoldshafen,~Germany}
\affiliation{PHI,~Karlsruhe~Institute~of~Technology,~76131~Karlsruhe,~Germany}
\author{D.~Rieger}
\affiliation{PHI,~Karlsruhe~Institute~of~Technology,~76131~Karlsruhe,~Germany}
\author{P.~Winkel}
\affiliation{PHI,~Karlsruhe~Institute~of~Technology,~76131~Karlsruhe,~Germany}
\author{F.~Henriques}
\affiliation{PHI,~Karlsruhe~Institute~of~Technology,~76131~Karlsruhe,~Germany}
\author{F.~Valenti}
\affiliation{PHI,~Karlsruhe~Institute~of~Technology,~76131~Karlsruhe,~Germany}
\affiliation{IPE,~Karlsruhe~Institute~of~Technology,~76344~Eggenstein-Leopoldshafen,~Germany}
\author{A.~Ionita}
\affiliation{PHI,~Karlsruhe~Institute~of~Technology,~76131~Karlsruhe,~Germany}
\author{M.~Wessbecher}
\affiliation{PHI,~Karlsruhe~Institute~of~Technology,~76131~Karlsruhe,~Germany}
\author{M.~Spiecker}
\affiliation{PHI,~Karlsruhe~Institute~of~Technology,~76131~Karlsruhe,~Germany}
\author{D.~Gusenkova}
\affiliation{PHI,~Karlsruhe~Institute~of~Technology,~76131~Karlsruhe,~Germany}
\author{I.~M.~Pop}
\affiliation{IQMT,~Karlsruhe~Institute~of~Technology,~76344~Eggenstein-Leopoldshafen,~Germany}
\affiliation{PHI,~Karlsruhe~Institute~of~Technology,~76131~Karlsruhe,~Germany}
\author{W.~Wernsdorfer}
\email[]{wolfgang.wernsdorfer@kit.edu}
\affiliation{IQMT,~Karlsruhe~Institute~of~Technology,~76344~Eggenstein-Leopoldshafen,~Germany}
\affiliation{PHI,~Karlsruhe~Institute~of~Technology,~76131~Karlsruhe,~Germany}

\date{\today}

\begin{abstract}
\noindent
High kinetic inductance materials constitute a valuable resource for superconducting quantum circuits and hybrid architectures. Superconducting granular aluminum (grAl) reaches kinetic sheet inductances in the $\si{\nano\henry}/\square$ range, with proven applicability in superconducting quantum bits and microwave detectors. Here we show that the single photon internal quality factor $Q_{\mathrm{i}}$ of grAl microwave resonators exceeds $10^5$ in magnetic fields up to \SI{1}{\tesla}, aligned in-plane to the grAl films.
Small perpendicular magnetic fields, in the range of $\SI{0.5}{\milli\tesla}$, enhance $Q_{\mathrm{i}}$ by 
approximately \SI{15}{\percent}, possibly due to the introduction of quasiparticle traps in the form of fluxons. Further increasing the perpendicular field deteriorates the resonators' quality factor. These results open the door for the use of high kinetic inductance grAl structures in circuit quantum electrodynamics and hybrid architectures with magnetic field requirements.
\end{abstract}

\keywords{Optical and microwave phenomena, Microwave Resonators, Superconducting RF, High-kinetic inductance, Granular Aluminum}
\maketitle

Thanks to their intrinsically low losses, superconducting materials are at the heart of
quantum information hardware\cite{Manucharyan2009, Pop2014, Lin2018, Earnest2018}, hybrid semiconducting-superconducting systems\cite{deLange2015, Larsen2015, Casparis2018}, kinetic inductance detectors\cite{Day2003} and magnetometers\cite{Clarke2005, Clarke2006}. High kinetic inductance superconductors are 
particularly appealing, because they enable
the fabrication of compact, high impedance circuit elements operating in the \si{\giga\hertz} range. Notable examples are Josephson junction arrays\cite{Manucharyan2012, Bell2012, Masluk2012},
NbN\cite{Grabovskij2008, Luomahaara2014, Zollitsch2019, Niepce2019}, NbTiN\cite{Samkharadze2016, Hazard2019, Kroll2019}, TiN\cite{Vissers2010, Leduc2010, Shearrow2018}, W\cite{Basset2019}, InO\cite{Dupre2017} and granular aluminum (grAl)\cite{Rotzinger2016, Gruenhaupt2018, Maleeva2018}.
Here we focus on grAl, which has already demonstrated internal quality factors in excess of $10^5$ in 
the single photon regime, while simultaneously packing high inductances in the $\si{\nano\henry}/\square$ range\citep{Gruenhaupt2018}. 
In moderate magnetic fields, up to a few \si{\milli\tesla}, grAl is currently being employed for fluxonium qubit superinductors \cite{Gruenhaupt2019}, kinetic inductance detectors\cite{Valenti2019, Henriques2019}, and as a source of non-linearity for transmon qubits\cite{Winkel2019}. However,  
the implementation of circuit quantum electrodynamics in hybrid systems requires superconducting resonators resilient to Tesla magnetic fields\cite{Samkharadze2016, Bienfait2016, Kroll2019, Xu2019}. In this letter, we demonstrate that grAl resonators with kinetic inductance exceeding \SI{1}{\nano\henry}/$\square$ maintain internal quality factors above $10^5$ under in-plane magnetic fields up to \SI{1}{\tesla}.

Granular aluminum is distinct from atomically disordered superconductors, as it consists of pure, crystalline Al clusters 
with an average diameter of 3~-~\SI{5}{\nano\meter} embedded in a matrix of amorphous AlO$_x$\cite{Cohen1968, Deutscher1973, Rotzinger2016}. 
This material can be modeled as a network of Josephson junctions\cite{Maleeva2018}, in which the effective Josephson energy can be tuned by adjusting the partial oxygen pressure during e-beam evaporation of pure aluminum. Using optical lithography on a c-plane sapphire substrate, we fabricated superconducting $\lambda/2$ grAl resonators similar to the ones used in Refs.~\cite{Gruenhaupt2018, Henriques2019}. The grAl film thickness is \SI{20}{\nano\meter} with a sheet resistivity in the 1.4~-~\SI{1.8}{\kilo\ohm}/$\square$ range corresponding to a kinetic inductance in the 1.2~-~\SI{1.5}{\nano\henry}/$\square$ range (cf.~Table~\ref{tab:summary}) and a critical temperature of $\approx~\SI{2}{\kelvin}$ (cf. Ref.~\cite{LevyBertrand2019}).

\begin{figure*}[htb]
  \center{\includegraphics[width=\textwidth]
        {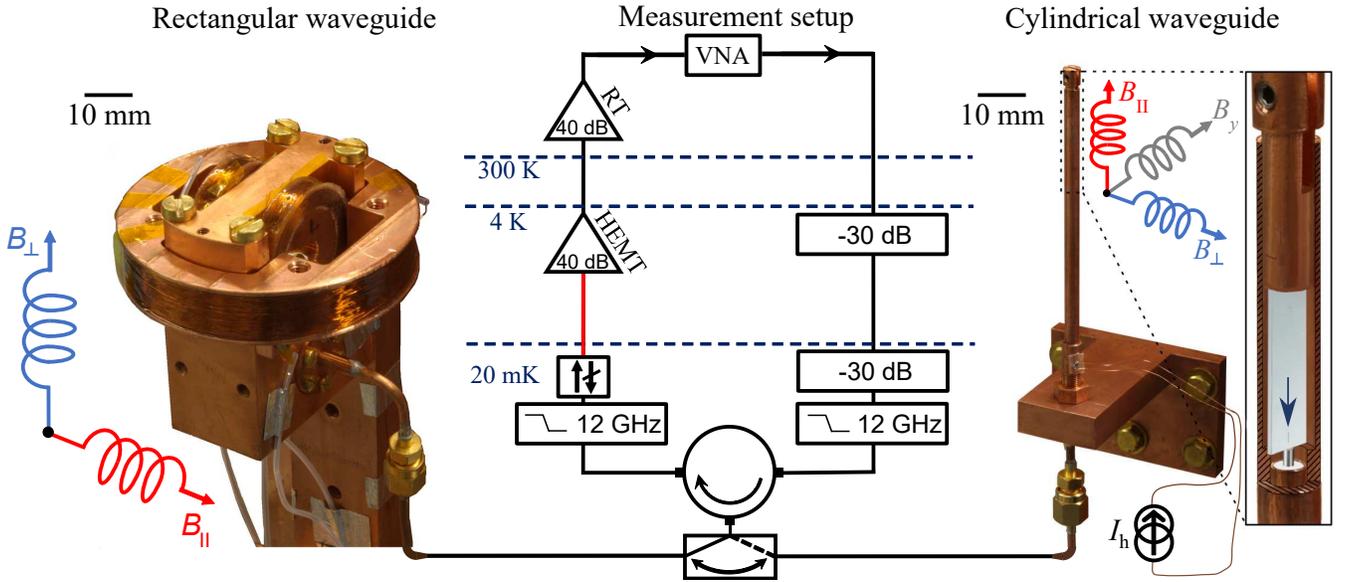}}
  \caption{Microwave reflection measurement setup connected either to a rectangular waveguide with 2D magnetic field control, or to a cylindrical waveguide with 3D field control. All experiments are performed at the base temperature ($\approx~\SI{20}{\milli\kelvin}$) of a table-top, Sionludi inverted dilution refrigerator\cite{Sionludi12}. In the middle, we present a schematic of the microwave measurement setup. The input line contains two \SI{30}{\decibel} attenuators at \SI{4}{\kelvin} and \SI{20}{\milli\kelvin}, a \SI{12}{\giga\hertz} low-pass filter and a circulator for signal routing, thermalized at \SI{20}{\milli\kelvin}. The output line contains a \SI{12}{\giga\hertz} low-pass filter, an isolator, connected through a superconducting Nb-Ti coaxial line (in red) to a low-noise high-electron mobility transistor (HEMT) amplifier at \SI{4}{\kelvin}, and a low-noise room temperature (RT) amplifier at \SI{300}{\kelvin}. The measurements are performed with a commercial Vector Network Analyser (VNA).   
Left side: Rectangular waveguide housing the samples, similarly to Refs.\cite{Kou2018, Gruenhaupt2018}. The in-plane field along the resonator axis, $B_{\parallel}$, is provided by two Helmholtz coils, and the perpendicular field, $B_{\perp}$, is provided by a single axial coil around the waveguide. All coils are thermalized to the mixing chamber. The maximum value for $B_{\parallel}$ is \SI{650}{\milli\tesla}. 
Right side: Cylindrical waveguide with full 3D magnetic field control, provided by a set of coils (not shown), thermalized at the \SI{4}{\kelvin} stage. In this configuration the maximum value for $B_{\parallel}$ is \SI{1.2}{\tesla}. By applying an approximately \SI{1}{\second} long current pulse $I_{\mathrm{h}}$ through a local heater attached to the cylindrical waveguide, the temperature of the resonator can be increased above $T_{\mathrm{C}}~\approx~\SI{2}{\kelvin}$ in order to reset persistent currents trapped in the grAl film. The resonator thermalizes below \SI{100}{\milli\kelvin} within \SI{5}{\minute} following the heating pulse. The zoom-in represents schematically the sapphire chip on which a $\lambda/2$ grAl resonator, at the position indicated by the blue arrow, is capacitively coupled to a coaxial cable.}
  \label{fig:setup}
\end{figure*} 

In Fig.~\ref{fig:setup} we show the two sample holders used to test the magnetic field resilience of superconducting grAl resonators.
Following the approach of Refs.~\cite{Paik2011, Kou2018},
the resonators are enclosed in 3D copper waveguides in order to reduce the surface dielectric participation and the associated radio-frequency dissipation. The waveguides are anchored to the mixing 
chamber of an inverted, table-top dilution refrigerator Sionludi\cite{Sionludi12} with a base temperature of \SI{20}{\milli\kelvin}. The rectangular waveguide\cite{Gruenhaupt2018} (left image in Fig.~\ref{fig:setup}) accommodates a pair of Helmholtz coils for in-plane field along the resonator's axis, $B_{\parallel}$ up to \SI{650}{\milli\tesla}, and a single coil for perpendicular field, $B_{\perp}$. To increase the maximum attainable magnetic field and to provide tri-axial field control, we designed a cylindrical waveguide holder (right image in Fig.~\ref{fig:setup}) with a remarkably small outer diameter of \SI{3.6}{\milli\meter}. Thanks to its reduced dimensions, the cylindrical waveguide thermalized at the dilution stage can be placed in a compact coil assembly with 3D field control up to $B^{\mathrm{max}}_{\parallel}=\SI{1.2}{\tesla}$, thermalized at the \SI{4}{\kelvin} stage. We perform reflection measurements using the microwave setup schematized in Fig.~\ref{fig:setup} (center). 

\begin{figure*}[htb]
\vspace{-1mm}
  \center{\includegraphics[width=\textwidth]
        {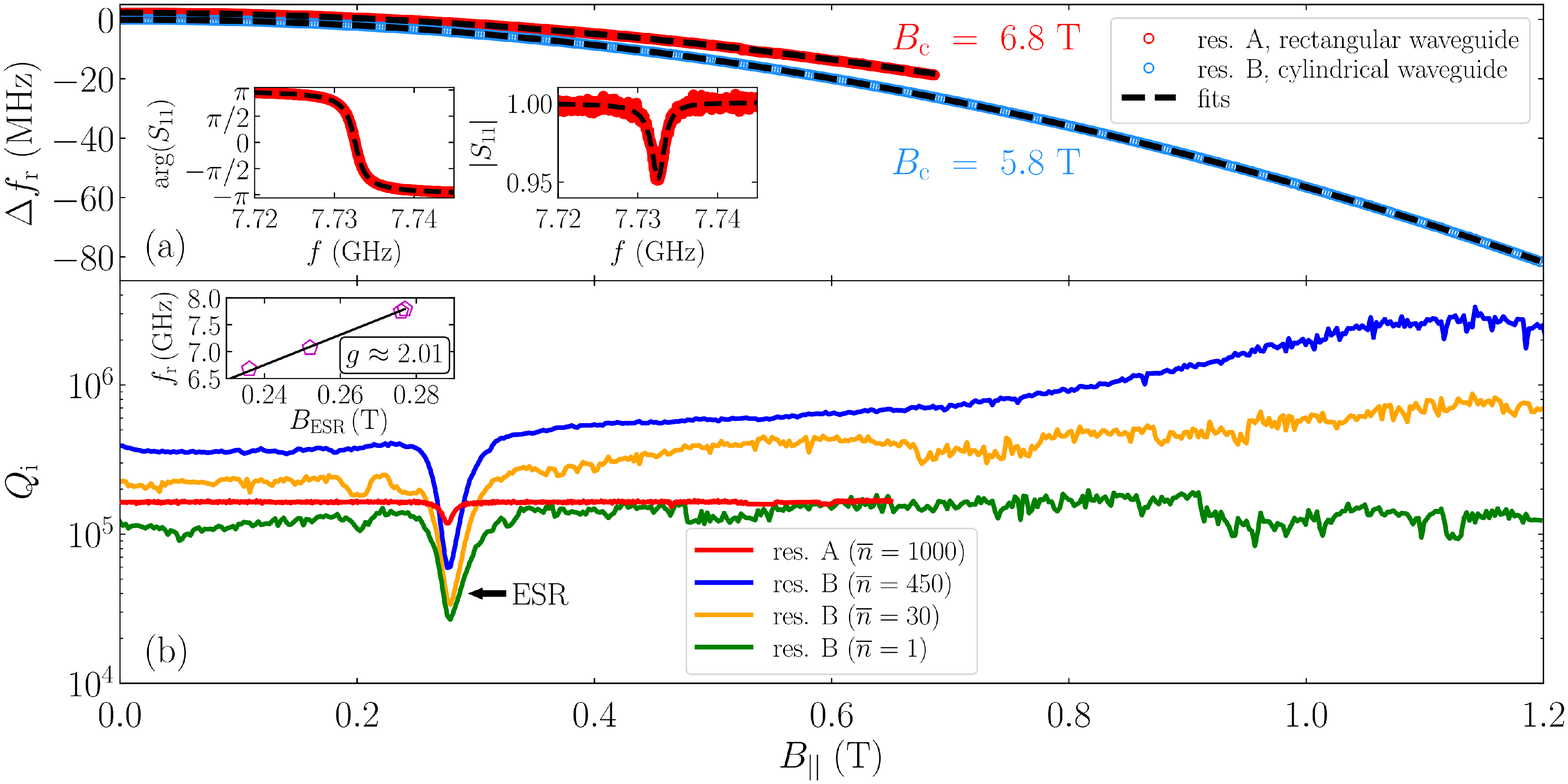}}
  \caption{Resonance frequency shift $\Delta f_{\mathrm{r}}$ and internal quality factor $Q_{\mathrm{i}}$ of grAl resonators versus in-plane magnetic field $B_{\parallel}$. For each $B_{\parallel}$ value, the spurious perpendicular component, resulting from field misalignment and inhomogeneity, is minimized by choosing the compensation field $B_{\perp}$  which maximizes the resonant frequency (cf. Fig.~{\ref{fig:field_alignment}}).
In panel (a), we present the measured $\Delta f_{\mathrm{r}}$ as a function of $B_{\parallel}$ for res. A (in red), in the rectangular waveguide, and for res. B (in blue), in the cylindrical waveguide, along with the corresponding fits following Eq.~\ref{eq:delta_fr} (dashed black). We extract a grAl critical field of \SI{6.8}{\tesla} and \SI{5.8}{\tesla} for res. A and B, respectively (cf. Table~\ref{tab:summary}). The two insets show typically measured phase and amplitude (in red) for res. A, at $B_{\parallel}~=~\SI{650}{\milli\tesla}$, and the corresponding joint fit~\cite{Probst2015} (in black).
In panel (b), we plot the fitted $Q_{\mathrm{i}}$ as a function of $B_{\parallel}$, for various circulating photon numbers, $\overline{n}$, populating the resonator.
Overall, we do not observe a degradation of $Q_{\mathrm{i}}$ with increasing $B_{\parallel}$.
Interestingly, the $Q_{\mathrm{i}}$ for res. B enhances versus in-plane field when populated with tens to hundreds of photons (yellow and blue markers). The conspicuous dips in $Q_{\mathrm{i}}$, indicated by the black arrow, can be explained by coupling to the electron spin resonance (ESR) of paramagnetic impurities\cite{Samkharadze2016, Kroll2019}. The inset shows the resonance frequencies of four different samples (in magenta) versus the measured ESR magnetic field extracted from the position of the $Q_{\mathrm{i}}$ dip. From a linear fit (in black), the Land\'{e} factor is $g~\approx~2.01$, which suggests coupling to a spin 1/2 ensemble.}
  \label{fig:fr_Qi_para}
\end{figure*}

The measured frequency shift of the grAl resonators with 
increasing in-plane field is shown in Fig.~\ref{fig:fr_Qi_para} (a). 
The participation ratio of the grAl kinetic inductance, $L_{\mathrm{k}}$, 
is close to unity\cite{Gruenhaupt2018}, therefore the fundamental mode frequency is $2\pi f_{\mathrm{r}}~\approx~1/\sqrt{L_{\mathrm{k}}C}$, where $C$ is the microstrip capacitance.
Following Mattis-Bardeen theory\cite{Mattis1958, Annunziata2010} for superconductors in the local and dirty limit, the total kinetic inductance is $L_{\mathrm{k}}(B_{\parallel})~=~N\hbar R_{\mathrm{sq}}/(\pi\Delta(B_{\parallel})~\times~\tanh(\Delta(B_{\parallel})/2k_{\mathrm{B}}T))$ where $N$ is the number of squares, $R_{\mathrm{sq}}$ is the resistance per square and $\Delta$ is the superconducting gap. Using the field dependence of the superconducting gap\cite{Douglass1961}, 
$\Delta(B_{\parallel})/\Delta(0)~=~\sqrt{1 - (B_{\parallel}/B_{\mathrm{c}})^2}$, where $B_{\mathrm{c}}$ is the critical field, the frequency shift can be approximated by:
\begin{equation}\label{eq:delta_fr}
\frac{\Delta f_{\mathrm{r}} (B_{\parallel})}{f_{\mathrm{r}}(B_{\parallel} = 0)} \approx  - \frac{1}{4}\left(\frac{B_{\parallel}}{B_{\mathrm{c}}}\right)^2.
\end{equation}
Using Eq.~\ref{eq:delta_fr}, we fit the measured frequency shift (cf. Fig.~\ref{fig:fr_Qi_para} (a)) and extract the critical field $B_{\mathrm{c}}$ for our grAl films in the range of
\SI{4.9}{\tesla}-\SI{6.8}{\tesla} (cf. Table~\ref{tab:summary}) consistent with previous measurements\cite{Abeles1967, Chui1981}. For res. D in the cylindrical waveguide, we measure a similar dependence of $\Delta f_{\mathrm{r}}$ versus $B_{\mathrm{y}}$ (cf. Fig.~\ref{fig:B_para_comp}), confirming that $\Delta f_{\mathrm{r}}$ is independent of the direction of the in-plane field (cf. Eq.~\ref{eq:delta_fr}). 

The resilience of grAl resonators to in-plane magnetic field $B_{\parallel}$ is demonstrated in Fig.~\ref{fig:fr_Qi_para}~(b), where we plot the internal quality factor $Q_{\mathrm{i}}$ as a function of $B_{\parallel}$. Resonators in both waveguide setups (cf.~Fig.~\ref{fig:setup}) maintain
$Q_{\mathrm{i}}~>~10^{5}$ up to $B_{\parallel}~=~\SI{650}{\milli\tesla}$ (rectangular, red) and $B_{\parallel}~=~\SI{1.2}{\tesla}$ (cylindrical, blue). 
The measured $Q_{\mathrm{i}}$ increases for higher number of circulating photons $\overline{n}$, as indicated by the green, yellow and blue traces corresponding to $\overline{n}~=~1,~45,~\mathrm{and}~100$. 
As proposed in Refs.~\cite{LevensonFalk2014, Gruenhaupt2018}, this dependence suggests circulating current in the resonator can accelerate quasiparticle diffusion. Interestingly, at fields in the range of \SI{1}{\tesla}, the power dependence of $Q_{\mathrm{i}}$ is approximately 6 $\times$ stronger than in zero field. This effect might be explained by imperfect spatial compensation of the perpendicular field, introducing vortices which can act as quasiparticle traps\cite{Nsanzineza2014}. As expected from Ref.\citep{Maleeva2018}, the self-Kerr frequency shift of the resonator vs. $\overline{n}$ is not influenced by $B_{\parallel}$ (cf. Fig.~\ref{fig:self_Kerr}).

The quality factor versus $B_{\parallel}$ shows a characteristic dip 
for all measured resonators (cf. Fig.~\ref{fig:fr_Qi_para} (b) for res. A and B), which can be attributed to electron spin resonance (ESR) of paramagnetic impurities\cite{Samkharadze2016, Kroll2019}.  The inset of Fig.~\ref{fig:fr_Qi_para}~(b) shows the frequency of the resonators vs. the magnetic field $B_{\mathrm{ESR}}$ at which the dip is observed. Using the resonance condition with the Zeeman-splitting energy 
$hf_{\mathrm{r}}~=~g\mu_{\mathrm{B}}B_{\mathrm{ESR}}$, we extract a Land\'{e} factor $g~\approx~2.01$ (cf. black line in inset). This points to a spin-1/2 ensemble of unknown origin coupled to the resonators. Following Ref.~\cite{Yang2019}, in the case of grAl the ensemble could consist of spins localized in the oxide between aluminum grains.

\begin{figure*}[htb]
\vspace{-6mm}
  \center{\includegraphics[width=\textwidth]
        {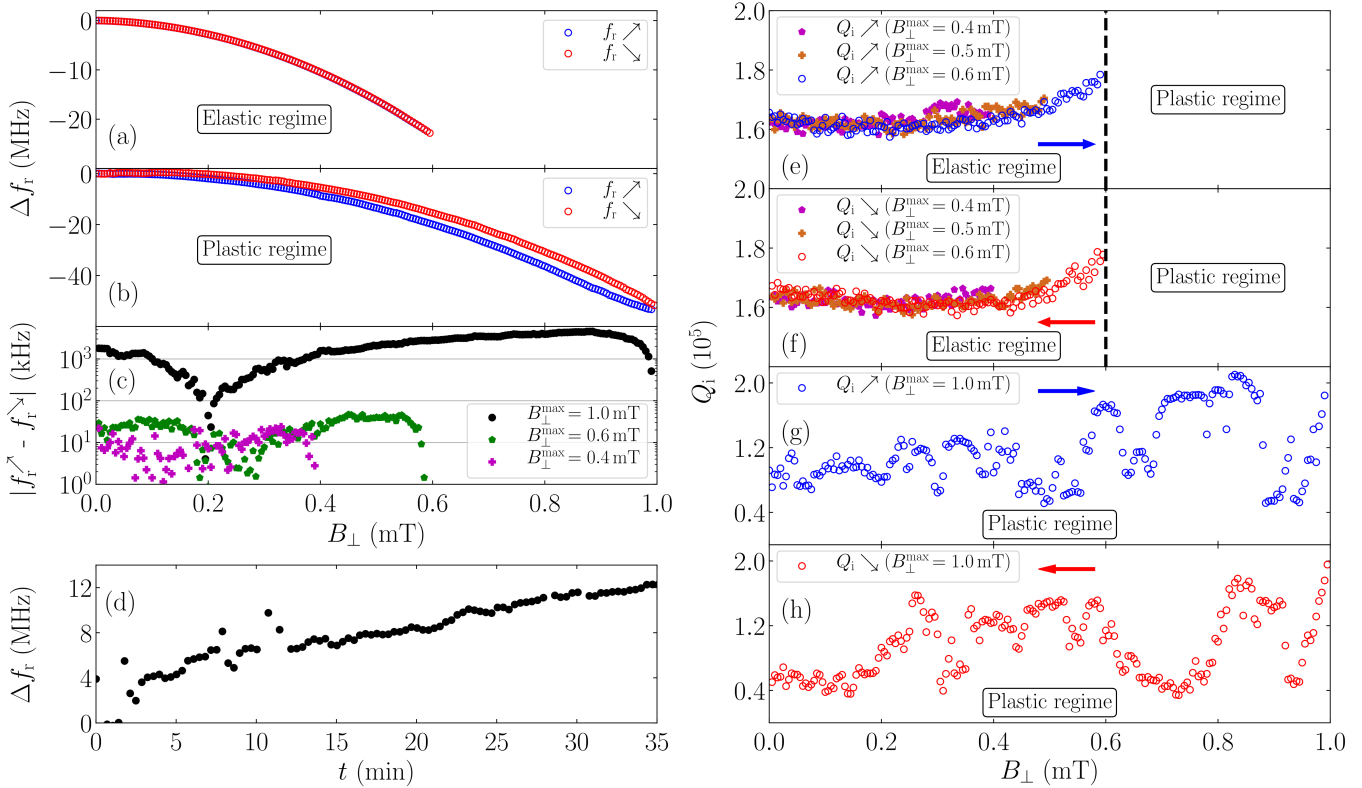}}
  \caption{Behaviour of grAl resonator A in perpendicular magnetic field. 
In panel (a), we show that the resonant frequency shift is reversible for $B_{\perp}^{\mathrm{max}}$ up to \SI{0.6}{\milli\tesla}. We denote this the \textit{elastic regime}. In contrast, panels (b) and (c) show that above this threshold the frequencies on the field ramp-up ($f_{\mathrm{r}}^{\nearrow}$) and ramp-down ($f_{\mathrm{r}}^{\searrow}$) do not overlap, indicating flux trapping in the resonator film. We denote this the \textit{plastic regime}. In the plastic regime, as shown in panel (d), we observe flux creeping on long time scales: the resonance frequency gradually drifts upwards.
We show $Q_{\mathrm{i}}(B_{\perp})$ in the elastic regime, measured during ramp-up (e) and ramp-down (f) for  $B_{\perp}^{\mathrm{max}}$ of \SI{0.4}{\milli\tesla}, \SI{0.5}{\milli\tesla} and \SI{0.6}{\milli\tesla}. Blue and red arrows denote the field sweep directions. Note that $Q_{\mathrm{i}}$ is enhanced by approximately $\SI{15}{\percent}$ in $B_{\perp}~=~\SI{0.6}{\milli\tesla}$. After having transitioned to the plastic regime (during the sweeps to $B_{\perp}^{\mathrm{max}} = \SI{0.8}{\milli\tesla}$ and $B_{\perp}^{\mathrm{max}} = \SI{0.9}{\milli\tesla}$ shown in Fig.~\ref{fig:onset_plastic}), the measured $Q_{\mathrm{i}}(B_{\perp})$ changes randomly both on the ramp-up (g) and ramp-down (h) for $B_{\perp}^{\mathrm{max}}~=~\SI{1.0}{\milli\tesla}$. 
}
  \label{fig:fr_Qi_perp}
\end{figure*} 

In Fig.~\ref{fig:fr_Qi_perp} we show the resonance frequency shift and the quality factor in perpendicular field. The field is successively swept to $B_{\perp}^{\mathrm{max}}$ and back to $0$, with $B_{\perp}^{\mathrm{max}}$ gradually increased. The aim is to determine the threshold perpendicular field, 
$B_{\mathrm{th}}$, beyond which the resonance frequencies on the ramp up and ramp 
down no longer coincide, due to flux trapping\cite{Stan2004}. 
A sample is in the so-called \textit{elastic regime} if it has not been exposed to fields above $B_{\mathrm{th}}$ after crossing the superconducting transition in zero field (cf.~Fig.~\ref{fig:fr_Qi_perp}~(a)). 
Notice that although the frequency shift is qualitatively similar to the one measured in $B_{\parallel}$ (cf. Fig.~\ref{fig:fr_Qi_para} (a)), the magnetic field susceptibility is 3 orders of 
magnitude stronger due to the larger area exposed to the field and the induced persistent currents. 
Once a sample is exposed to fields larger than $B_{\mathrm{th}}$, it enters a so-called \textit{plastic regime}, defined by randomly pinned and mobile fluxons with varying configurations versus $B_{\perp}$ (cf.~Fig.~\ref{fig:fr_Qi_perp}~(b) and (c)). For our resonators we measure $B_{\mathrm{th}}~\approx~\SI{0.6}{\milli\tesla}$. 
After applying $B_{\perp}~=~\SI{2}{\milli\tesla}$, deep in the plastic regime, and ramping down to zero, we observe an upward drift of the resonance frequency in time (cf.~Fig.~\ref{fig:fr_Qi_perp}~(d)). This trend can be attributed to mobile fluxons exiting the film.
From the plastic regime, a sample can be reset to the elastic regime by heating it above $T_{\mathrm{c}}$ and cooling it down in zero field. This reset is achieved in about \SI{5}{\minute} utilizing the local heater visible in ~Fig.~\ref{fig:setup} (on the cylindrical waveguide).

\begin{table}
\begin{center}
\begin{tabular}{ |c||c|c|c|c|c|  }
 \hline
 Resonator & Waveguide & $Q_{\mathrm{c}}$ & $f_{\mathrm{r}}$ & $L_{\mathrm{k}}$ & $B_{\mathrm{c}}$\\
  & & & (GHz) & ($\si{\nano\henry}/\square$) & (T)
\\ 
 \hline
 A & Rectangular & 3.0 x 10$^3$ & 7.75 & 1.2 & 6.8\\
 B & Cylindrical & 1.1 x 10$^4$ & 7.79 & 1.2 & 5.8\\
 C & Rectangular & 1.7 x 10$^4$ & 6.68 & 1.4 & 6.0\\
 D & Cylindrical & 1.8 x 10$^6$ & 7.07 & 1.5 & 4.9\\
 \hline
\end{tabular}
\caption{Summary of the coupling quality factor $Q_{\mathrm{c}}$, resonance frequency $f_{\mathrm{r}}$, kinetic sheet inductance $L_{\mathrm{k}}$ and fitted critical field $B_{\mathrm{c}}$ (cf.~Eq.~\ref{eq:delta_fr} and Fig.~\ref{fig:fr_Qi_para}~(a)). The main source of uncertainty in estimating $B_{\mathrm{c}}$ originates in the frequency shift caused by inhomogeneities in the spurious out-of-plane field.}
\label{tab:summary}
\end{center}
\vspace{-6mm}
\end{table}

In the elastic regime, the internal quality factor improves by approximately \SI{15}{\percent} in perpendicular field of \SI{0.6}{\milli\tesla} (cf. Fig.~\ref{fig:fr_Qi_perp} (e) and (f)), which can be explained by fluxons created at the current nodes\cite{Nsanzineza2014}. The $Q_{\mathrm{i}}$ enhancement 
disappears when $B_{\perp}~\rightarrow~0$, which indicates that fluxons are induced by reversible circulating currents. 
In the plastic regime, $Q_{\mathrm{i}}$ changes randomly with $B_{\perp}$, due to fluxons interacting with the radio-frequency current of the resonator mode (cf.~Fig.~\ref{fig:fr_Qi_perp} (g) and (h)).
The onset of the plastic regime, evidenced by a sharp drop in $Q_{\mathrm{i}}$, occurs during the first sweep exceeding $B_{\perp}^{\mathrm{max}}~=~\SI{0.7}{\milli\tesla}$, shown in Fig.~\ref{fig:onset_plastic}.

In summary, we demonstrated that superconducting grAl resonators maintain internal 
quality factors above $10^5$ under in-plane fields exceeding \SI{1}{\tesla}. The observed 
enhancement of the $Q_{\mathrm{i}}$ in small perpendicular fields reinforces the notion that 
fluxons, trapped at particular positions, can facilitate quasiparticle relaxation. 
Above a perpendicular field threshold, trapped fluxons lead to drifts and stochastic jumps of the resonator's frequency and quality factor. This threshold is geometry dependent, and for the \SI{10}{\micro\meter}-wide resonators used in this work is in the range of \SI{0.6}{\milli\tesla}.
To further decrease the susceptibility to perpendicular fields, the width of the resonator should be decreased 
in future designs. The ease of fabrication, the kinetic inductance in the $\si{\nano\henry}/\square$ range and 
the field resilience up to \SI{1}{\tesla} recommend grAl as a material for hybrid quantum systems. 

We acknowledge fruitful discussions with H. Rotzinger, U. Vool and W.
Wulfhekel. We thank S. Diewald, L. Radtke and the KIT Nanostructure Service Laboratory for technical support. Funding was provided by the Alexander von Humboldt foundation in the framework of a Sofja
Kovalevskaja award endowed by the German Federal Ministry of Education and Research, and by the Initiative and Networking Fund of the Helmholtz Association, within the Helmholtz Future Project Scalable solid state quantum computing. KB, DR, PW and WW acknowledge support from the European Research Council advanced grant MoQuOS (N. 741276).

The data that support the findings of this study are available from the corresponding author upon reasonable request.

\bibliography{grAl_field}

\newpage 
\renewcommand{\thefigure}{S\arabic{figure}}
\renewcommand{\thesection}{S\arabic{section}}
\setcounter{figure}{0}

\onecolumngrid
\newpage
\section*{Supplementary Material}
\hrulefill

\vspace{4 cm}

\twocolumngrid

\section{Minimization of the perpendicular field component}
\begin{figure}[htb]
  \center{\includegraphics[width=\columnwidth]
        {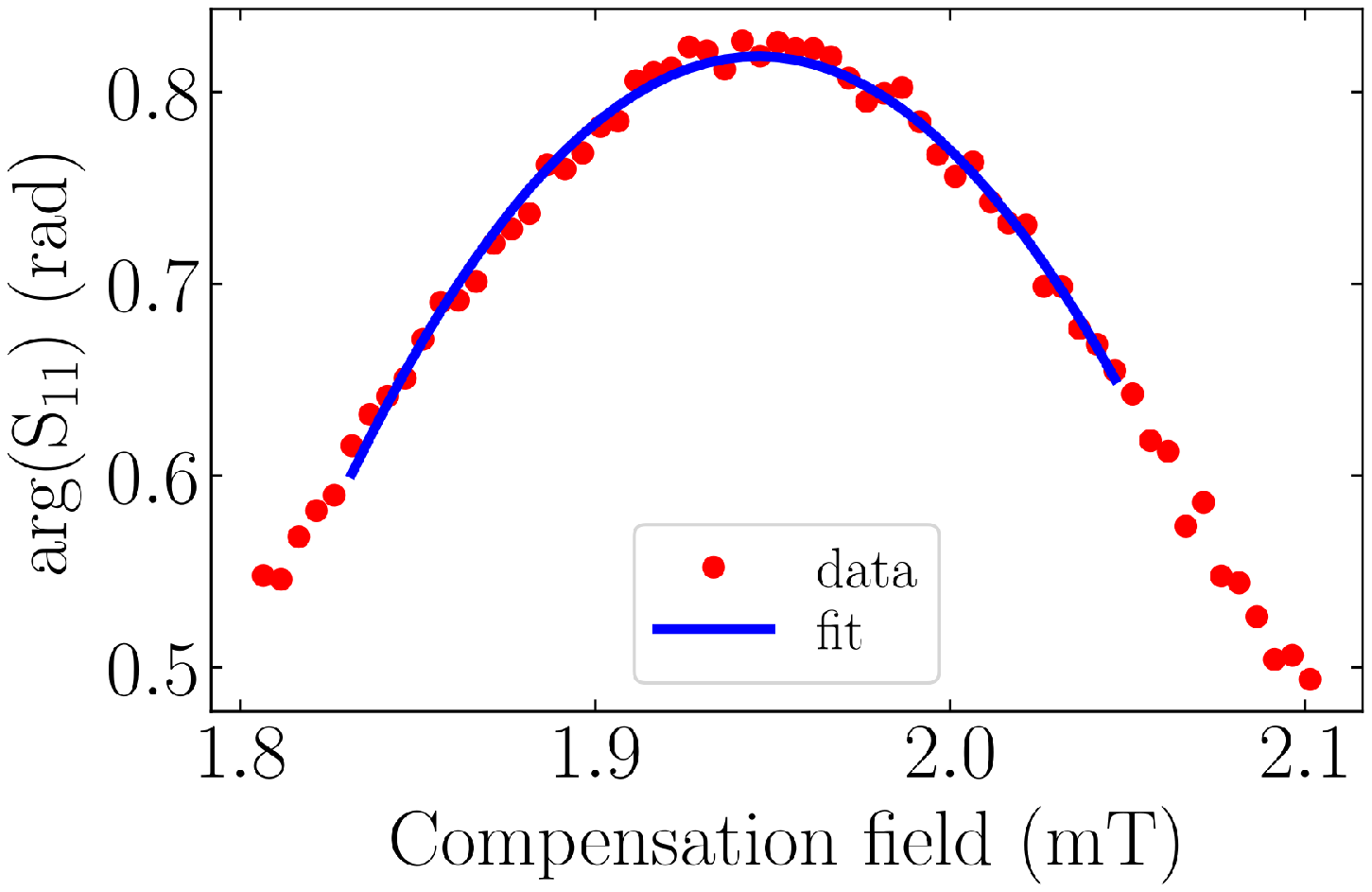}}
  \caption{Typical measurement used to calibrate the compensation field $B_{\perp}$. The phase response of the resonator at a fixed frequency close to $f_{\mathrm{r}}$ is shown in red and the quadratic fit in blue. We choose the value of the compensation field which maximizes the phase response. For the example shown here, res. A and $B_{\parallel}~=~\SI{0.2}{\tesla}$, the compensation field is $B_{\perp}~\approx~\SI{1.94}{\milli\tesla}$.}
  \label{fig:field_alignment}
\end{figure}

As evidenced in Fig.~\ref{fig:fr_Qi_perp} (a) and (b), $B_{\perp}$ shifts strongly the 
resonance frequency. We use this susceptibility to minimize the spurious perpendicular field 
component during in-plane sweeps. For each value of $B_{\parallel}$, we trace the phase 
response of the resonator at a fixed frequency (close to $f_{\mathrm{r}}$) in a narrow field range of $B_{\perp}$ (cf. Fig.~\ref{fig:field_alignment}). The maximum of the phase response, corresponding to the optimal compensating $B_{\perp}$, is determined from quadratic fit. We observe a compensation field linearly-dependent on the in-plane magnetic field, hence, a minor chip tilt with respect to Helmholtz coils is the most probable origin for the unwanted perpendicular field component. A typical compensation field of $\approx~\SI{1.94}{\milli\tesla}$ required for $B_{\parallel}~=~\SI{200}{\milli\tesla}$ (cf. Fig.~\ref{fig:field_alignment}) corresponds to a misalignment angle below \SI{1}{\degree}.

\section{In-plane field sweeps along orthogonal axes}
\begin{figure}[htb]
  \center{\includegraphics[width=\columnwidth]
        {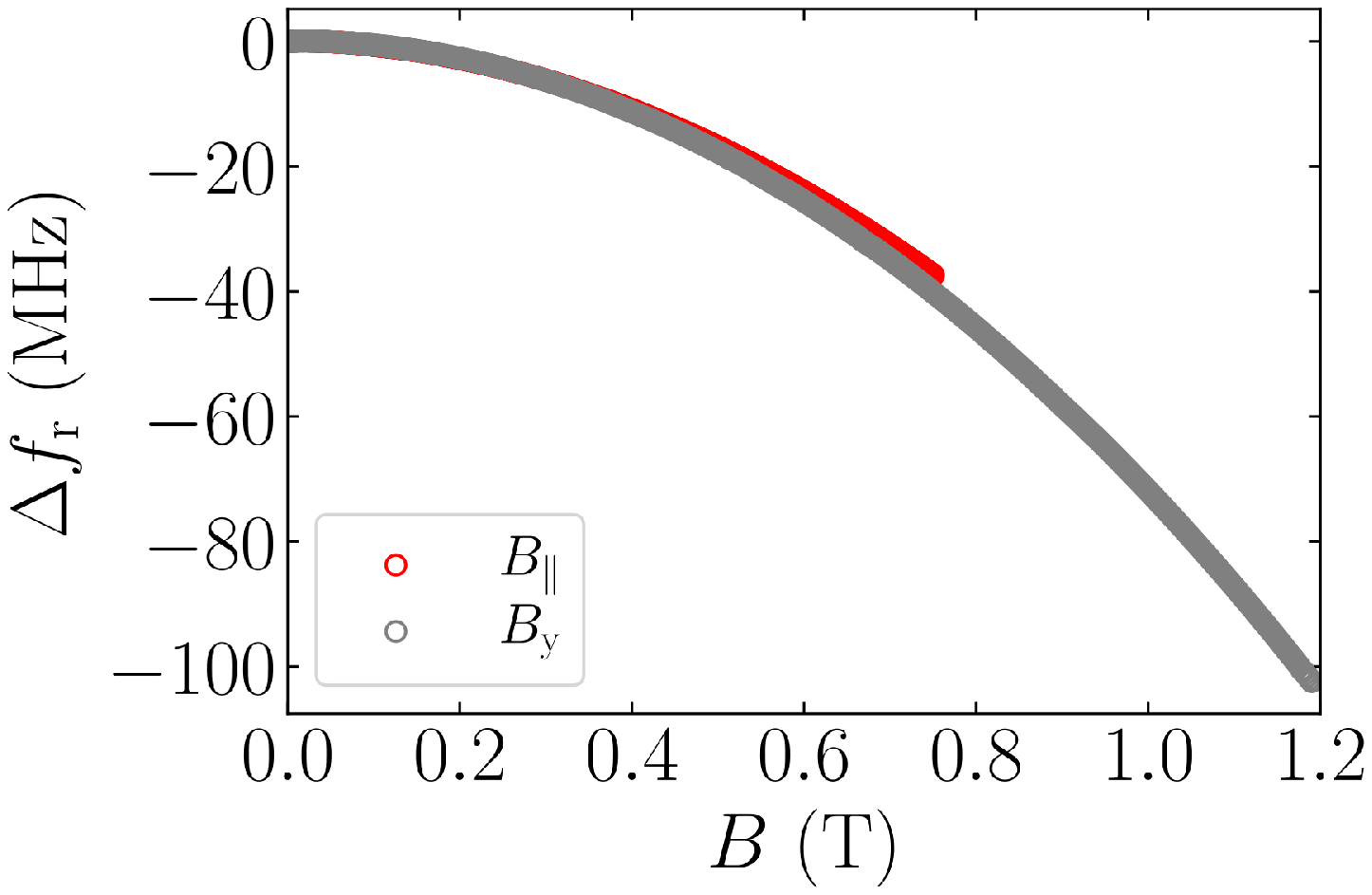}}
  \caption{Resonance frequency shift versus in-plane magnetic field for res. B (measured in the cylindrical waveguide), applied along two orthogonal directions: parallel (red) and perpendicular (gray) to the resonator's axis. The fact that the measurements overlap, for different effective side areas of the resonator exposed to in-plane field, indicates that screening currents do not play a significant role. Active compensation of the spurious perpendicular field component is performed for both sweeps, according to the procedure shown in Fig.~\ref{fig:field_alignment}.}
  \label{fig:B_para_comp}
\end{figure}

A comparison between the resonance frequency shift for two different in-plane magnetic field directions is shown for res. B in Fig.~\ref{fig:B_para_comp}. Magnetic field sweeps along the resonator's axis (red) and perpendicular to the resonator's axis (gray) overlap closely, which indicates that in our case the orientation of the in-plane field does not significantly influence the superconducting properties of the film. Since the effective areas of the resonator parallel and perpendicular to the its axis are 60 $\times$ different, the suppression of the superconductor's gap appears to be the main mechanism responsible for the measured change of kinetic inductance under in-plane magnetic field (cf. Eq.~\ref{eq:delta_fr} in the main text).  

\section{Self-Kerr effect vs. in-plane field}

\begin{figure}[hb!]
  \center{\includegraphics[width=\columnwidth]
        {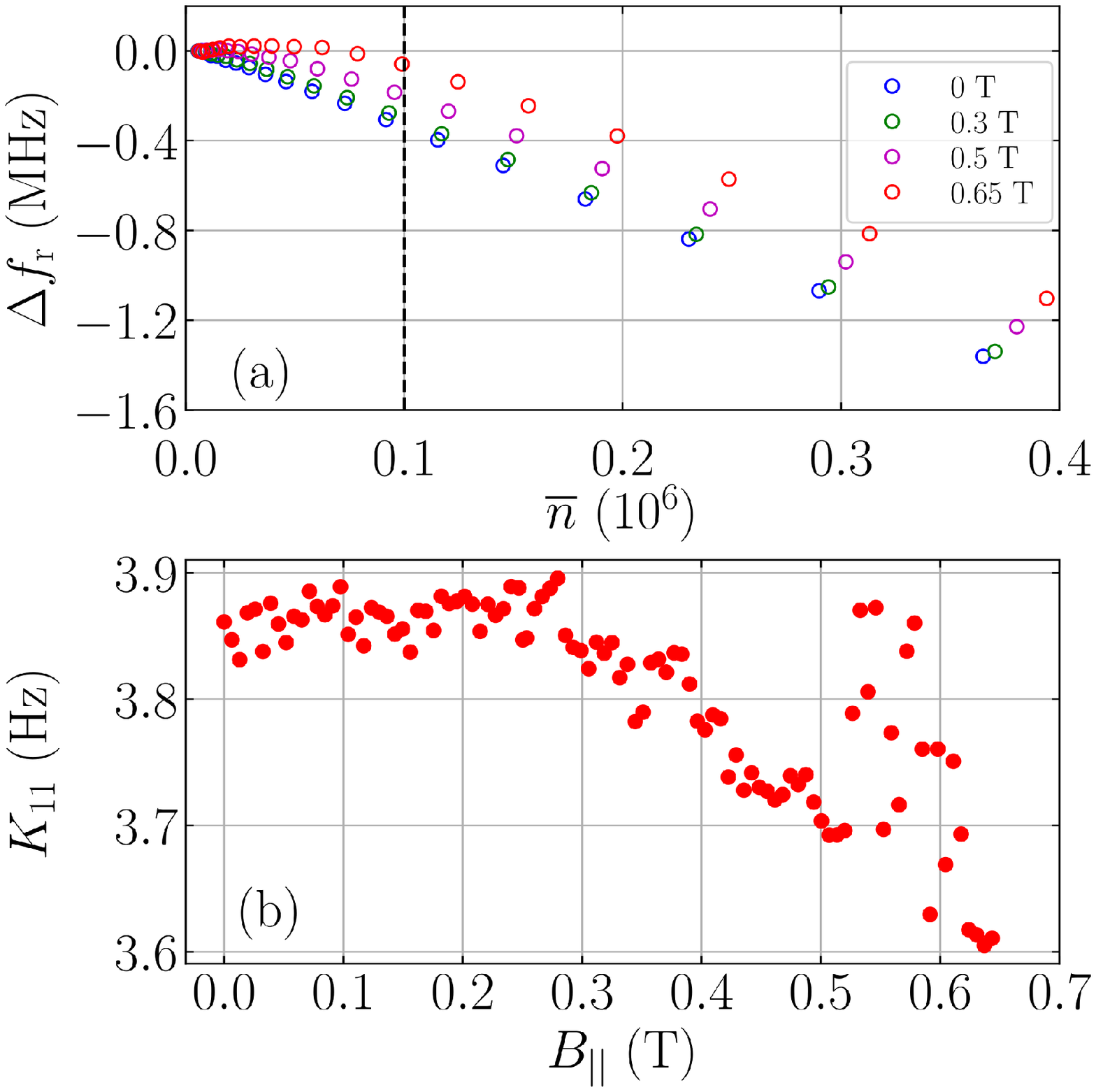}}
  \caption{Self-Kerr coefficient of res. A under in-plane field. (a) Resonance frequency shift vs. the average photon number, $\overline{n}$, for $B_{||}~=~\SI{0}{\tesla}, \SI{0.3}{\tesla}, \SI{0.5}{\tesla}, \SI{0.65}{\tesla}$. The self-Kerr coefficient is extracted from the data above $\overline{n}~\approx~1~\cdot~10^{5}$ (indicated by vertical black dashed line) using a linear fit, $\Delta f_{\mathrm{r}}~=~-K_{11}\overline{n}$. This resonator bifurcates at $\approx~4~\cdot~10^{5}$ photons. (b) Fitted self-Kerr coefficient vs. $B_{||}$. }
  \label{fig:self_Kerr}
\end{figure} 

The resonance frequency shift as a function of the number of circulating photons in the resonator, known as the self-Kerr effect, is measured under in-plane magnetic field for res. A (cf. Fig.~\ref{fig:self_Kerr}~(a)).
Note that for $B_{||}~>~\SI{0.5}{\tesla}$ the resonance frequency shift is no longer linear in the low photon number region. Furthermore at \SI{0.65}{\tesla}, $\Delta~f_{\mathrm{r}}$ first increases and then decreases. The exact cause of this behaviour is unclear. It can be attributed to induced superconducting fluxons, due to imperfect compensation of the perpendicular field in the end regions of the resonator, at the current nodes. These fluxons can act as quasiparticle traps\cite{Nsanzineza2014}, which become more efficient at higher circulating power in the resonator when the mobility of the quasiparticles is higher. Trapping quasiparticles into these areas could slightly reduce the effective quasiparticle density, $n_{\mathrm{qp}}$, and as $f_{\mathrm{r}}~\propto~1/\sqrt{L_{\mathrm{k}}}~\propto~1/\sqrt{n_{\mathrm{qp}}}$, result in an increased $f_{\mathrm{r}}$.

The extracted self-Kerr coefficient under in-plane magnetic field up to \SI{650}{\milli\tesla} is shown in Fig.~\ref{fig:self_Kerr}~(b). As discussed in Ref.~\cite{Maleeva2018}, the self-Kerr coefficient of grAl is $K_{11}~\propto~f_{\mathrm{r}}^2/j_{\mathrm{c}}$, where $f_{\mathrm{r}}$ and $j_{\mathrm{c}}$ are the resonant frequency and the critical current density. Because both $f_{\mathrm{r}}^{2}$ and $j_{\mathrm{c}}$ are $\propto~1/L_{\mathrm{k}}$, $K_{11}$ is expected to be field-independent\cite{Maleeva2018}.

\section{Onset of the plastic regime under perpendicular magnetic field}

\begin{figure}[hb!]
  \center{\includegraphics[width=\columnwidth]
        {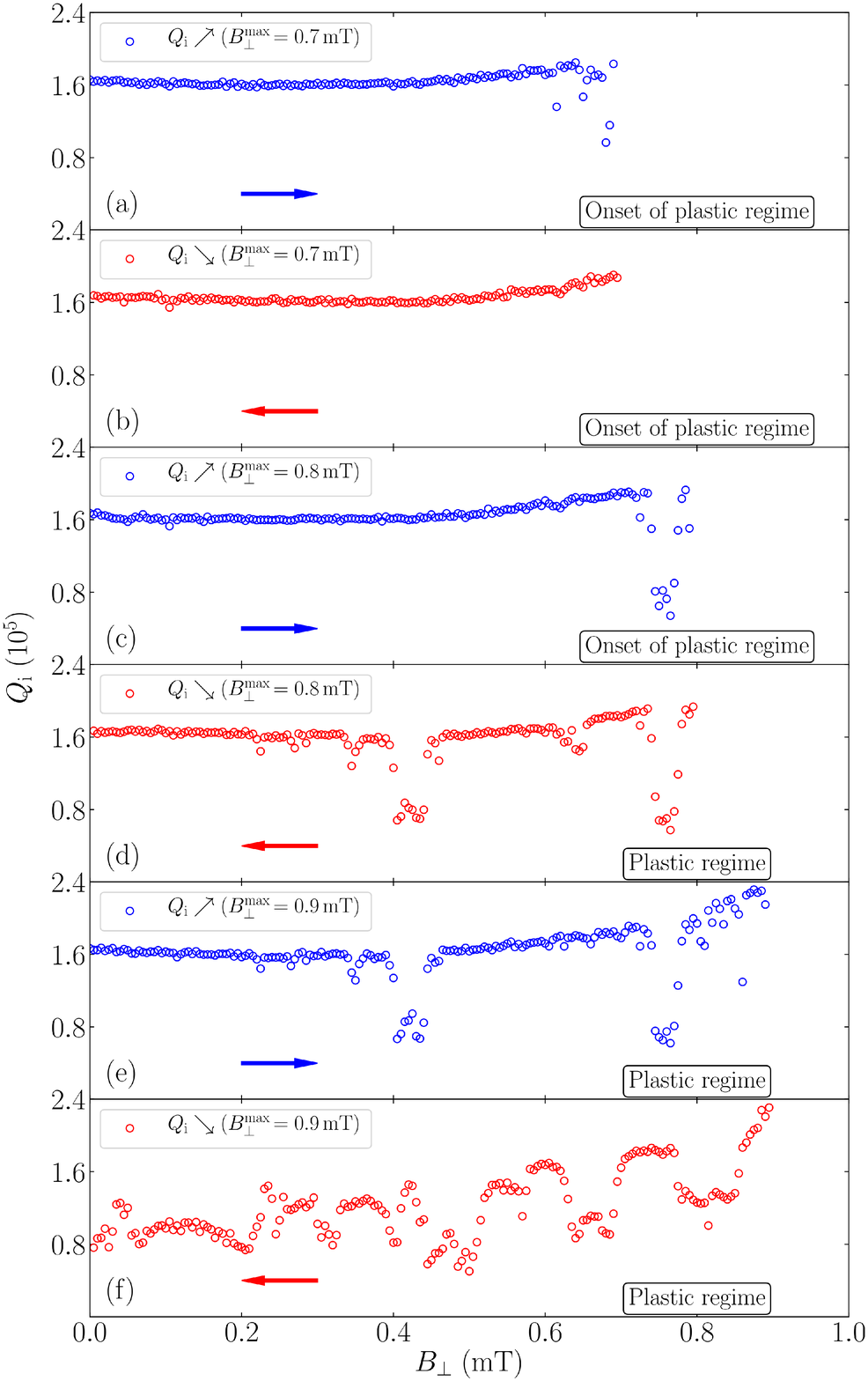}}
  \caption{Measurements of $Q_{\mathrm{i}}$ versus perpendicular magnetic field for res. A, evidencing the onset of the plastic regime during the sweeps to $B_{\perp}^{\mathrm{max}} = \SI{0.7}{\milli\tesla}$ (a - b), $B_{\perp}^{\mathrm{max}} = \SI{0.8}{\milli\tesla}$ (c - d) and $B_{\perp}^{\mathrm{max}} = \SI{0.9}{\milli\tesla}$ (e - f). Horizontal blue and red arrows indicate the directions of the magnetic field sweep. 
}
  \label{fig:onset_plastic}
\end{figure}

The transition of resonator A to the plastic regime (cf. Fig.~\ref{fig:fr_Qi_perp}) is visible during three consecutive sweeps with $B_{\perp}^{\mathrm{max}} = \SI{0.7}{\milli\tesla}, \SI{0.8}{\milli\tesla}~\mathrm{and}~\SI{0.9}{\milli\tesla}$ (cf. Fig.~\ref{fig:onset_plastic}). When $B_{\perp}$ is ramped up to 
\SI{0.8}{\milli\tesla}, the onset of the plastic regime is evidenced by a sharp drop in 
the $Q_{\mathrm{i}}$ for $B_{\perp}~>~\SI{0.7}{\milli\tesla}$, which is explained by fluxons permeating the film (cf. Fig.~\ref{fig:onset_plastic} (c)).

\end{document}